\documentclass{iopart}
\usepackage{graphicx}
\usepackage{wasysym}

\begin{document}

\title[Spin liquids: a large-$S$ route]
{Quantum spin liquids: a large-$S$ route.}

\author{Oleg Tchernyshyov}

\address{Department of Physics and Astronomy, Johns Hopkins University,
3400 N. Charles St., Baltimore, MD 21218, USA}

\begin{abstract}
This paper explores the large-$S$ route to quantum disorder in the
Heisenberg antiferromagnet on the pyrochlore lattice and its
homologues in lower dimensions.  It is shown that zero-point
fluctuations of spins shape up a valence-bond solid at low
temperatures for one two-dimensional lattice and a liquid with very
short-range valence-bond correlations for another.  A one-dimensional
model demonstrates potential significance of quantum interference
effects (as in Haldane's gap): the quantum melting of a valence-bond
order yields different valence-bond liquids for integer and
half-integer values of $S$.
\end{abstract}

\section{Introduction}

Frustrated magnets have recently become a focus of experimental
studies \cite{Schiffer96,Zinkin96,Greedan01}.  Frustration
disrupts long-range spin order---even at low temperatures---and leads
to the formation of a spin-liquid state, in which spins move in a
random yet strongly correlated fashion.  A very large degeneracy of
the classical ground state makes these magnets strongly susceptible to
a variety of perturbations \cite{Moessner01,Tch02b} and thereby leads
to a plethora of possible thermodynamic phases.  It is also thought
that the interplay of strong correlations and quantum effects may
yield quantum ground states without magnetic order.  Low-temperature
properties of such magnets will be quite distinct from those of the
familiar Bose or Fermi liquids.  Therefore quantum ground states and
low-energy excitations of frustrated magnets arouse considerable
interest.

\begin{figure}
\centerline{\includegraphics[width=0.7\columnwidth]{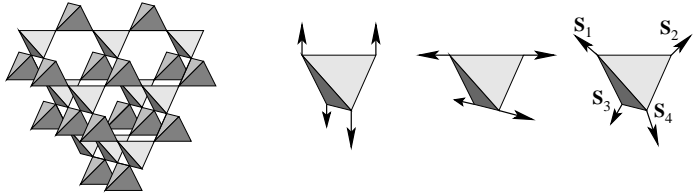}}
\caption{The pyrochlore lattice and sample ground states of Heisenberg
spins on a tetrahedron.}
\label{fig-pyrochlore}
\end{figure}

The ultimate example of strong frustration is the Heisenberg
antiferromagnet on the pyrochlore lattice \cite{Moessner98}.  In
magnets of this kind spins form a three-dimensional network of
corner-sharing tetrahedra (Fig.~\ref{fig-pyrochlore}).  A nearly
perfect realization of this model is found in ZnCr$_2$O$_4$
\cite{Lee02} where Cr$^{3+}$ ions have spin $S=3/2$.  Recent
theoretical studies of such systems have concentrated on the case
$S=1/2$ \cite{Harris91,Canals98,Koga01,Fouet03,Tsunetsugu02,Altman03},
for which the quantum effects are most prominent.

The purpose of this paper is to demonstrate viability of an
alternative approach: the large-$S$ route.  There are several reasons
to take this circuitous path.  First, the method provides a systematic
way to study quantum effects as a function of $S$.  Second, because
the very concept of a frustrated magnet is defined in the classical
context, staying close to the classical limit puts the discussion on
firmer ground.  Third, magnets of this kind have propensity toward
spin-Peierls order even in the classical limit \cite{Tch02b}.
Therefore, there is a reasonable chance of finding nonclassical
states---such as valence-bond solids and liquids---in pyrochlore
antiferromagnets at large $S$.

It must be realized that the large-$S$ problem is highly nontrivial:
the starting point ($S=\infty$) has an extensive degeneracy
\cite{Moessner98}.  The original problem for the three-dimensional
pyrochlore lattice remains unsolved.  Below I describe a few positive
results that have been obtained recently for homologues of the
pyrochlore antiferromagnet in one and two spatial dimensions.
Sec.~\ref{sec-perturbative} outlines a degenerate perturbation theory,
due to Henley, that is based on a series expansion in powers of $1/S$.
To the first order in $1/S$, the theory yields a valence-bond solid
for one two-dimensional ``pyrochlore'' and a valence-bond liquid for
another.  By design, the {\em ad hoc} series expansion misses effects
nonanalytical in $1/S$ (cf. Haldane's gap).
Sec.~\ref{sec-nonperturbative} describes such an effect in a
one-dimensional ``pyrochlore'' chain with a spin-Peierls ground state.
In that model, the quantum melting of a valence-bond solid produces
qualitatively different valence-bond liquids for integer and
half-integer spins.

\section{A $1/S$ expansion for antiferromagnets on pyrochlore-like lattices}
\label{sec-perturbative}

The starting point of the $1/S$ expansion is the limit $S \to \infty$.
The ground states are found by minimizing the exchange energy $J
\sum_{\langle ij \rangle} ({\bf S}_i \cdot {\bf S}_j) = \mathcal
O(S^2)$ with respect to classical Heisenberg spin variables $\{{\bf
S}_i\}$.  In the case of the pyrochlore antiferromagnet, the total
spin of every tetrahedron must vanish (Fig.~\ref{fig-pyrochlore}).  A
simple counting argument \cite{Moessner98} reveals that the manifold
of classical vacua is formidably large: it contains one continuous
degree of freedom per tetrahedron.

The next term in the $1/S$ expansion contains a quantum correction
coming from zero-point fluctuations of spin waves \cite{Henley89}:
\begin{equation}
E^{(1)} = {\rm const} + \sum_{a} \case{1}{2} \hbar \omega_a
= \mathcal O(S),
\label{eq-order-S}
\end{equation}
where $\{\omega_a\}$ are spin-wave frequencies.
Because magnon spectra are not the same in different ground states,
the quantum correction (\ref{eq-order-S}) lifts the classical
degeneracy.  Ordinarily, quantum fluctuations select a few collinear
ground states related to each other by symmetries of the lattice
\cite{Henley87}.  In some special cases, the residual degeneracy may
be quite large and not linked to any obvious lattice symmetry
\cite{Chalker91}.

The pyrochlore lattice presents such an exception: Henley has found
that an infinite number of collinear states are selected by zero-point
fluctuations \cite{Henley}.  These vacua are not related to each other
by lattice symmetries and most are not even periodic.  The exact
number of the degenerate ground states is not known.  This accidental
degeneracy has been linked by Henley to certain transformations that
leave the spin-wave spectrum $\{\omega_a\}$ invariant.  The
transformations consist of flipping every spin on a (potentially
infinite) subset of tetrahedra.

To determine which of the colinear ground states have the lowest
zero-point energy, Henley has expressed the sum over the spin-wave
frequencies (\ref{eq-order-S}) as the trace of an operator dependent
on the spin values in a given ground state.  The result can be
expanded as an infinite sum of multi-spin potentials.  In a collinear
spin configuration polarized along $\hat{\bf n}$, this interaction can
be expressed in terms of Ising variables $\sigma_i = ({\bf S}_i \cdot
\hat{\bf n})/S = \pm 1$, subject to the constraint
\begin{equation}
\textstyle{ \sum_{i=1}^4 } \  
\sigma_i=0 \hskip 3mm \mbox{ on every tetrahedron.}
\label{eq-constraint}  
\end{equation}
The zero-point energy reads \cite{Henley}
\begin{equation}
E^{(1)} = 
\case{3}{8}JS\sum_{\hexagon} 
\prod_{i=1}^{6} \sigma_i 
+ \ldots 
\hskip 5mm \mbox{(pyrochlore lattice).}
\label{eq-trunc-6}
\end{equation}
The six-spin interaction couples spins residing around regular
hexagons that exist in the pyrochlore lattice
(Fig.~\ref{fig-pyrochlore}).  The omitted terms, representing
multi-spin interactions of higher orders, are not necessarily small.
However, at least in the two-dimensional models discussed below, their
omission does not affect the selection of ground states.  One can
check, with the aid of Henley's symmetry, that all ground states of
the truncated Hamiltonian (\ref{eq-trunc-6}) have the same zero-point
energy (\ref{eq-order-S}).

\begin{figure}
\centerline{\includegraphics[width=0.75\columnwidth]{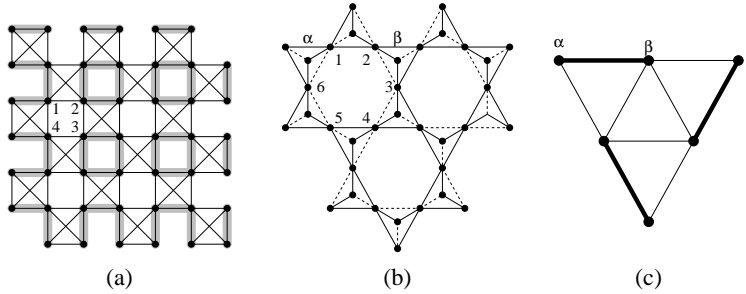}}
\caption{Two-dimensional lattices built from corner-sharing
tetrahedra: (a) The checkerboard.  Valence-bond order induced at low
temperatures ${\cal O}(JS/k_B)$ is shown.  $\langle {\bf S}_i \cdot
{\bf S}_j \rangle = -S^2$ for the shaded bonds and zero for the rest.
(b) One of the ground states of the pyrochlore wafer.  Dashed lines
designate frustrated bonds ($\tau_{ij} = -\sigma_i \sigma_j = -1$).
(c) The dimer state that corresponds to the bond state of one
sublattice of tetrahedra.}
\label{fig-obd}
\end{figure}

The problem of finding the ground states thus reduces to minimizing
the multi-spin interaction (\ref{eq-trunc-6}) over the Ising variables
$\{\sigma_i\}$, subject to the constraint (\ref{eq-constraint}).  It
remains unsolved for the pyrochlore lattice.  We have succeeded in
solving it for two other lattices built from corner-sharing
tetrahedra: the checkerboard and the pyrochlore wafer
(Fig.~\ref{fig-obd}).  The checkerboard is a projection of the
pyrochlore lattice onto a plane.  Because it has loops of length 4
(around empty squares), the zero-point potential starts with four-spin
interactions \cite{Henley,Tch03}:
\begin{equation}
E^{(1)} = 
-\case{1}{2}JS\sum_{\Square} 
\prod_{i=1}^{4} \sigma_i 
+ \ldots
\hskip 5mm \mbox{(checkerboard).}
\label{eq-trunc-4}
\end{equation}
For the pyrochlore wafer, a slice of the three-dimensional pyrochlore
lattice, the effective zero-point energy is given by
Eq.~\ref{eq-trunc-6}.

It is convenient to switch from {\em spin} variables $\sigma_i$ to
{\em bond} variables $\tau_{ij} = -\sigma_i \sigma_j = -({\bf S}_i
\cdot {\bf S}_j)/S^2$, defined for nearest-neighbor links $ij$.  In
terms of these, the constraint (\ref{eq-constraint}) requires that
exactly two non-adjacent bonds be frustrated ($\tau_{ij}=-1$) on every
tetrahedron.  The zero-point effective potentials acquire the
following form \cite{Tch03}:
\begin{equation}
-\prod_{i=1}^{4} \sigma_i = -\tau_{12} \tau_{34} = -\tau_{23} \tau_{41},
\hskip 5mm
\prod_{i=1}^{6} \sigma_i = -\tau_{12} \tau_{34} \tau_{56} 
= -\tau_{23} \tau_{45} \tau_{61}.
\label{eq-trunc-bond}
\end{equation}
Note that the lattices in question contain two sublattices of
tetrahedra and that the multi-spin interactions (\ref{eq-trunc-bond})
involve bonds of tetrahedra belonging to the same sublattice---see
Fig.~\ref{fig-obd}.  Thus the problem reduces to {\em independent}
minimizations of the bond energies (\ref{eq-trunc-bond}) on the two
tetrahedral sublattices.  Although not every bond configuration
$\{\tau_{ij}\}$ corresponds to a physical spin configuration
$\{\sigma_i\}$, those that minimize the bond potentials
(\ref{eq-trunc-bond}) do yield legitimate spin states \cite{Tch03}.

In the checkerboard case, the bond potential (\ref{eq-trunc-bond}) is
minimized when every two bonds facing each other across an empty
square [Fig.~\ref{fig-obd}(a)] are in the same state (e.g., $\tau_{12}
= \tau_{34} = +1$ and $\tau_{23} = \tau_{41} = -1$).  The ground
states of each tetrahedral sublattice fall into two disjoint classes.
In one class, which can be called ``horizontal stripes'', tetrahedra
of a given row are in the same bond state.  The frustrated bonds are
either all vertical, or all diagonal.  Because every row can be in one
of the two states, there are $2^{L/2}$ vacua for each sublattice.
The second class---vertical stripes---is related to the first by a
90-degree rotation of the (sub)lattice.  A more detailed derivation
will be given elsewhere \cite{Tch03}.

At zero temperature, the system is frozen in one of the striped ground
states breaking the rotational symmetry of the lattice.  There are two
Ising order parameters: each sublattice of tetrahedra independently
chooses to have vertical or horizontal stripes.  The discrete nature
of the order parameter guarantees survival of the long-range order at
low temperatures $T<T_c$.  The stripes will have a finite extent $\xi
\sim \mathcal O(\e^{a/T})$ enabling the system to move between ground
states with the same orientational order on a time scale of order
$\xi$.  Thermal expectation values for the bond variables $\langle
\tau_{ij} \rangle$ must therefore be averaged over all ground states
with a specified orientational order.  For example, when both
sublattices of tetrahedra have horizontal stripes, the horizontal
links always have antiparallel spins ($\langle \tau_{ij} \rangle =
+1$), while all other bonds have uncorrelated spins ($\langle
\tau_{ij} \rangle = 0$).  Another thermal state, with horizontal and
vertical stripes on the two sublattices of tetrahedra, is depicted in
Fig.~\ref{fig-obd}(a).  This state has antiparallel spins around one
half of the empty squares and strongly resembles the valence-bond
solid found in the ground state of the $S=1/2$ system \cite{Fouet03}.

Apart from the number of dimensions, the checkerboard and pyrochlore
lattices have one other important difference: on the checkerboard,
diagonal bonds are not equivalent to vertical or horizontal ones.
This is why the valence-bond order parameter on the checkerboard is of
the Ising, rather than $Z_3$, type.  The pyrochlore wafer
[Fig.~\ref{fig-obd}(b)] is free from this defect.  What classical
ground states are selected by quantum fluctuations there?

The 6-spin potential (\ref{eq-trunc-6}) translates into the
interaction (\ref{eq-trunc-bond}) involving three bonds around a
hexagon.  Again, the interaction takes place between tetrahedra of the
same sublattice.  The 3-bond energy is minimized when zero or two
bonds are frustrated ($\tau_{ij}=-1$).  While superficially this looks
like the ground-state rule for the checkerboard, the properties of the
ground states are entirely different.  In particular, there is no
valence-bond order.  Moreover, the connected valence-bond correlations
$\langle \tau_{ij} \tau_{kl} \rangle - \langle \tau_{ij} \rangle
\langle \tau_{kl} \rangle$ are extremely short-ranged.  This can be
proven by mapping the ground states of the bond potential
(\ref{eq-trunc-bond}) onto classical dimer coverings of the triangular
lattice, whose properties are well known \cite{Fendley02}.

The mapping is illustrated in Figs.~\ref{fig-obd}(b) and (c).  Treat
the tetrahedra of one sublattice as sites of a triangular lattice.
Every two tetrahedra contributing frustrated bonds to the same hexagon
(such as $\alpha$ and $\beta$) generate a dimer linking the
corresponding sites of the triangular lattice.  Because each
tetrahedron contributes exactly one frustrated bond to some hexagon,
every site of the dual lattice is connected by a dimer to another
site.

\section{Nonperturbative effects: the Berry phase}
\label{sec-nonperturbative}

\begin{figure}
\centerline{\includegraphics[width=0.33\columnwidth]{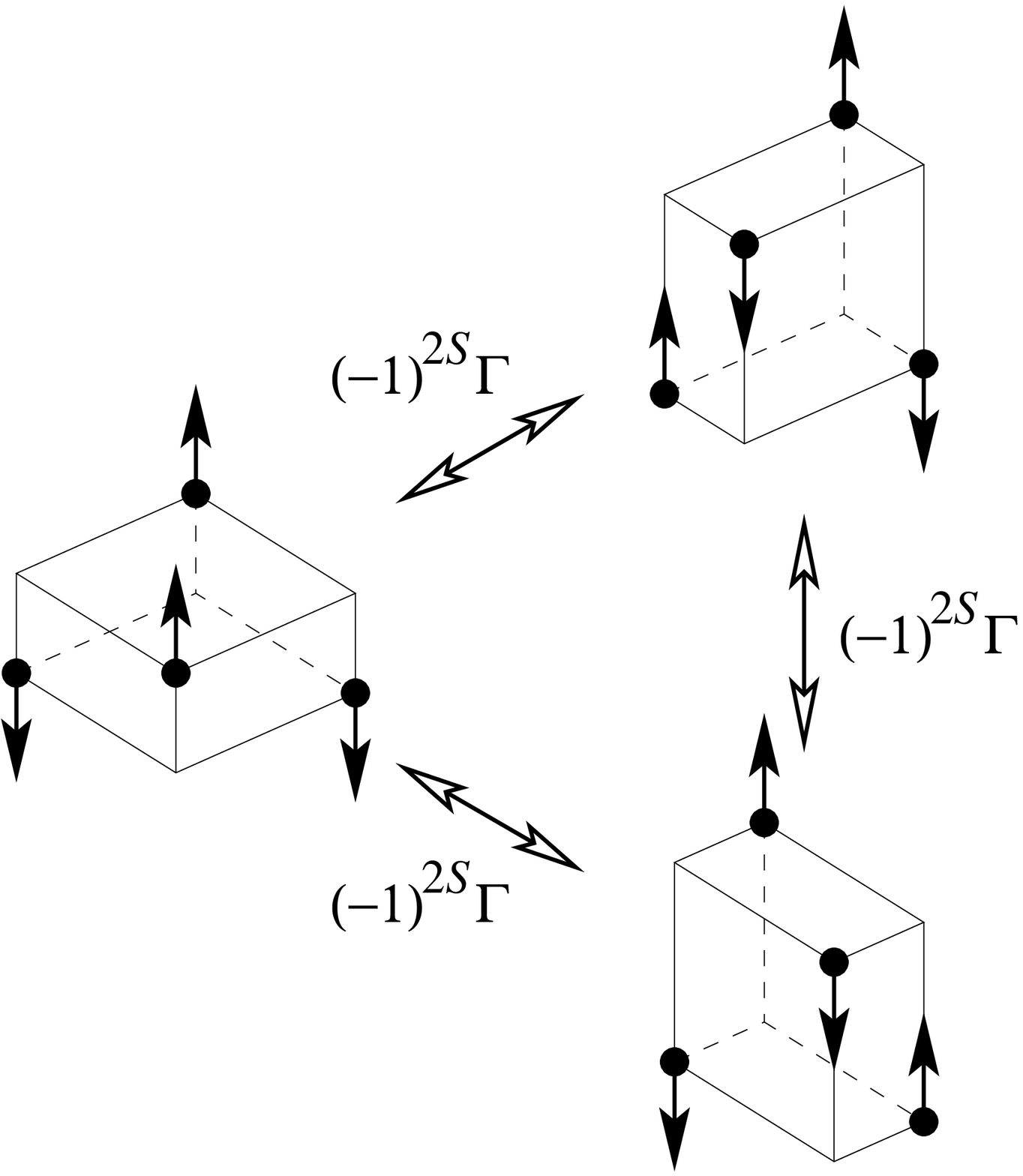}
\hspace{0.02\columnwidth}
\includegraphics[width=0.26\columnwidth]{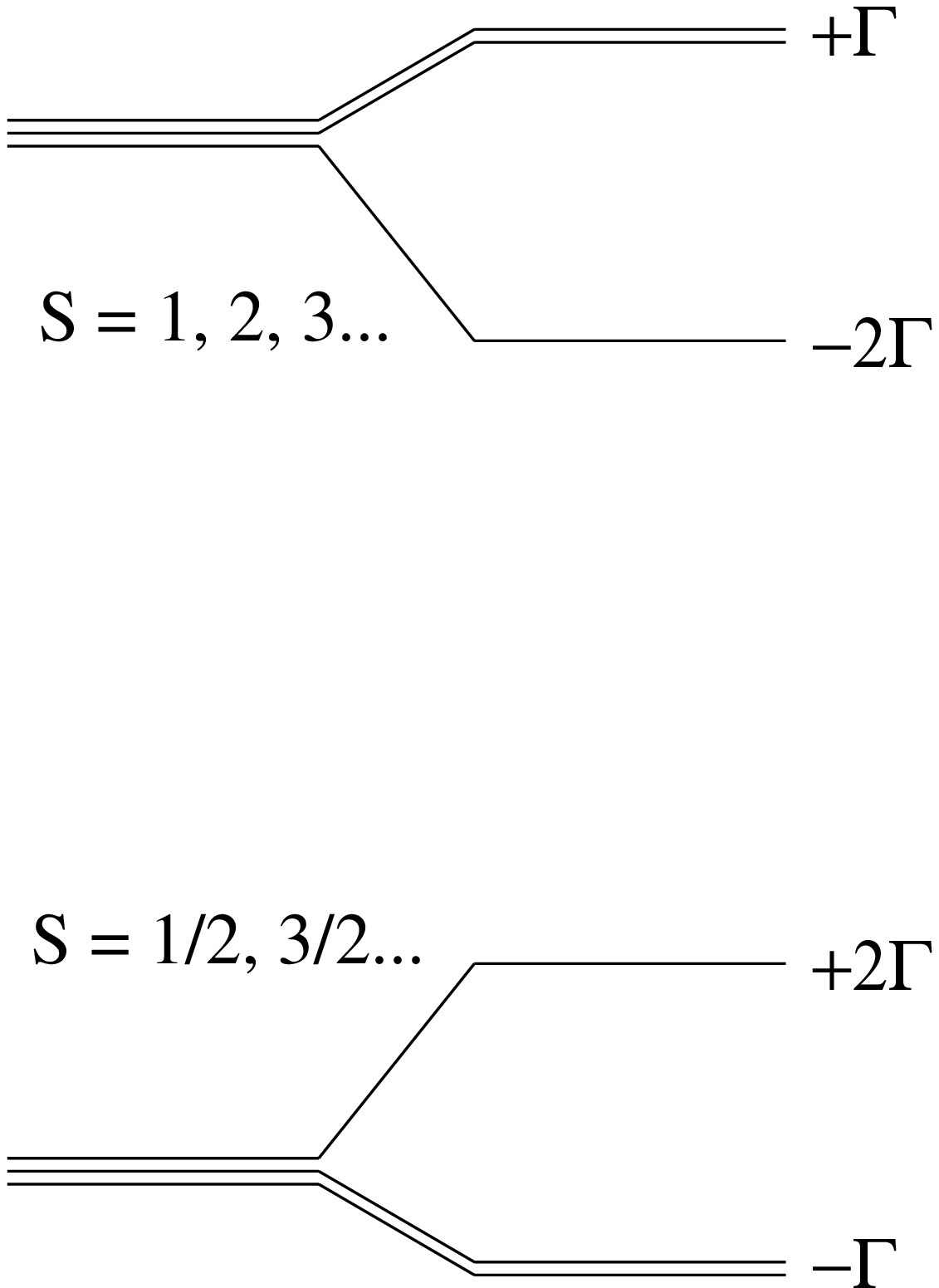}
\includegraphics[width=0.31\columnwidth]{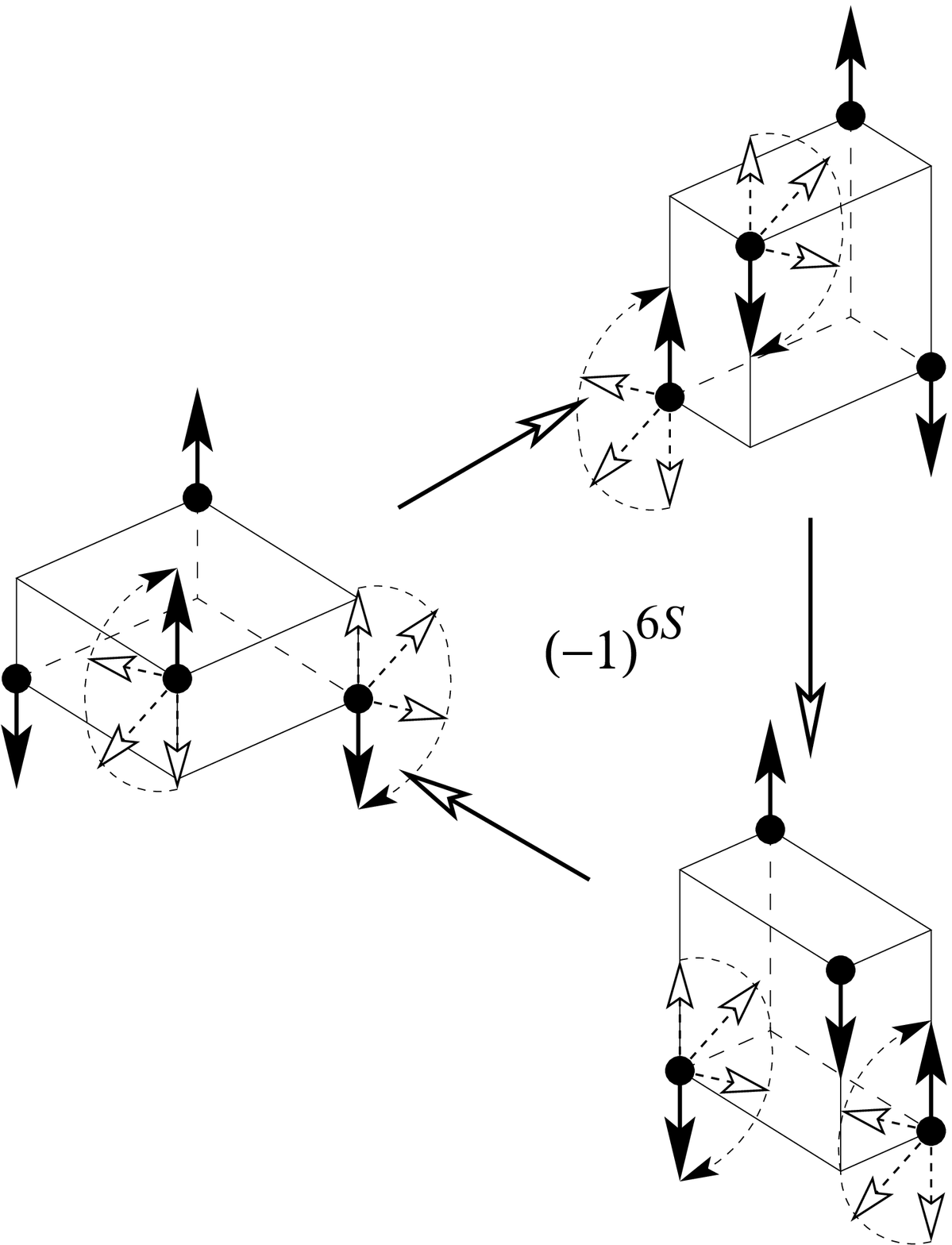} }
\caption{Tunneling lifts the degeneracy of the three distorted ground
states of spins on a tetrahedron.  The ground state is a singlet for
integer $S$ and a non-Kramers doublet for half-integer $S$.  The
difference can be traced to a Berry phase acquired by the spins in a
cycle of adiabatic evolution.}
\label{fig-DJT}
\end{figure}

To demonstrate potential significance of effects nonperturbative in
$1/S$, consider a toy model of four antiferromagnetically coupled
spins on a flexible tetrahedron.  On a regular tetrahedron, the ground
state is a spin singlet with the degeneracy $2S+1$.  A high degree of
symmetry and a non-Kramers degeneracy induce the Jahn--Teller effect.
For $S>1/2$, the sum of elastic and magnetic energies is minimized by
three distorted states.  In each of these the ``molecule'' is
flattened along one of its principal axes (left panel of
Fig.~\ref{fig-DJT}) \cite{Tch02b}.

Taking into account the kinetic energy of the atoms introduces
tunneling events between the three potential wells.  The tunneling
splits the degenerate ground states into a singlet and a doublet
(middle panel of Fig.~\ref{fig-DJT}).  The nontrivial result is an
oscillatory dependence of the order of the levels on $S$: the ground
state is a singlet for integer spins and a non-Kramers doublet for
half-integer spins.

At large $S$, the difference can be traced to a Berry phase acquired
by the spins as the system moves between the three distorted states
(the right panel of Fig.~\ref{fig-DJT}).  In the process, depicting
one of the potential tunneling paths, three spins make full $2\pi$
rotations in the same plane.  The overall amplitude thus acquires a
geometric phase $e^{6\pi iS}$, i.e. $+1$ if $S$ is integer and $-1$ if
it is half-integer.

\begin{figure}
\centerline{\includegraphics[width=0.8\columnwidth]{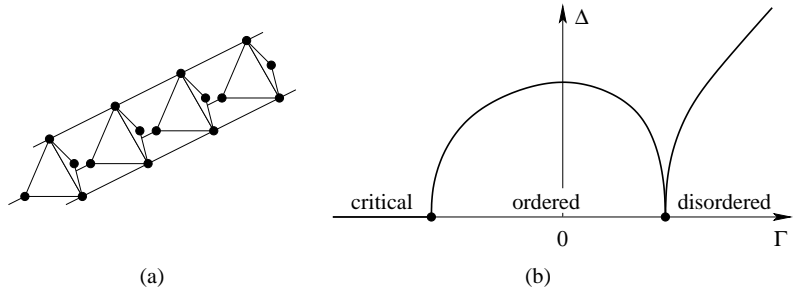}}
\caption{(a) A chain of weakly coupled tetrahedra with Heisenberg
spins.  For antiferromagnetic interactions, the low-energy sector
consists of singlet states. (b) The phase diagram of the $q=3$ quantum
Potts model showing the energy gap for valence-bond excitations
$\Delta$ as a function of the tunneling amplitude $\Gamma$.  The
region $\Gamma>0$ ($\Gamma<0$) describes the chain with integer
(half-integer) spins.  }
\label{fig-chain} 
\end{figure}

Consider now a generalization of this toy model to $d=1+1$ dimensions:
a chain of such tetrahedra weakly coupled through vibrational modes of
the lattice [Fig.~\ref{fig-chain}(a)].  The low-energy sector of the
model consists of three distorted spin-singlet ground states on each
site and can be described by the quantum $q=3$ Potts model.  Elastic
interactions cause the distortions of the tetrahedra to be correlated
with one another.  Assume, for the sake of the argument, that a
uniform distortion of the chain is favored.  Then, in the absence of
tunneling, the chain has spin-Peierls order with a $Z_3$
ferrodistortive order.  Low-energy spin-singlet excitations are domain
walls with a finite energy gap $\Delta$.  As the tunneling $\Gamma$ is
turned on, the spin-Peierls order weakens.  The singlet energy gap
vanishes at a quantum critical point [Fig.~\ref{fig-chain}(b)].  The
nature of the quantum-disordered phase depends on the sign of the
tunneling amplitude $\Gamma$ and thus on whether $S$ is integer or
half-integer.

The physics of low-energy singlet states is described by the quantum
3-state Potts model with the Hamiltonian \cite{Howes83}
\begin{equation}
H = - 2\sum_{i} \left[ 
\cos{\left(\case{2\pi}{3} (\theta_{i+1} - \theta_i)\right)}
+ \Gamma \cos{\left(\case{2\pi}{3} p_i\right)} 
\right]. 
\end{equation}  
Here $\theta_i = 0, 1, 2$ is the Potts variable, while $p_i = 0, 1, 2$
is its conjugate momentum.  On the $\Gamma>0$ side, the
quantum-disordered state also has a gap to singlet excitations.  For
$\Gamma<0$, the disordered state is gapless.

It is thus seen that the quantum melting of a valence-bond solid on
the chain proceeds in different ways for integer and half-integer
spins.  The quantum-disorderd phase has gapped singlet excitations
when $S$ is integer.  For half-integer $S$, the liquid has gapless
excitations in the form of valence-bond fluctuations.

\end{document}